\documentclass{PoS}

\title{Probing Interstellar Silicate Dust Grain Properties in Quasar Absorption Systems at Redshifts z$<$1.4}

\ShortTitle{Probing Interstellar Silicate Dust Grain Properties in Quasar Absorption Systems at Redshifts z$<$1.4}

\author{\speaker{Monique C. Aller}$^a$, 
Varsha P. Kulkarni$^a$, Donald G. York$^b$, Daniel E. Welty$^b$, Giovanni Vladilo$^c$, and Debopam Som$^a$ \\
\llap{$^a$} University of South Carolina, Department of Physics and Astronomy, Columbia, SC, USA\\
\llap{$^b$} University of Chicago, Department of Astronomy \& Astrophysics, Chicago, IL, USA\\
\llap{$^c$} Osservatorio Astronomico di Trieste, Via Tiepolo 11, 34143 Trieste, Italy\\
        E-mail: \email{moniquealler@gmail.com}, \email{kulkarni@sc.edu}, \email{don@oddjob.uchicago.edu}, \email{dwelty@oddjob.uchicago.edu}, \email{vladilo@oats.inaf.it}, \email{som@physics.sc.edu}}

\abstract{Absorption lines in the spectra of distant quasars whose sightlines serendipitously pass through foreground galaxies provide a valuable tool to simultaneously probe the dust and gas compositions of the interstellar medium (ISM) in galaxies. In particular, the damped and sub-damped Lyman-$\alpha$ (DLA/sub-DLA) absorbers trace gas-rich galaxies, independent of the intrinsic luminosities or star-formation rates of the associated galaxy stellar populations. The first evidence of silicate dust in a quasar absorption system was provided through our detection of the 10~$\mu$m silicate feature in the z=0.52 DLA absorber toward the quasar AO 0235+164. We present results from 2 follow-up programs using archival Spitzer Space Telescope infrared spectra to study the interstellar silicate dust grain properties in a total of 13 quasar absorption systems at $0.1<z<1.4$. We find clear detections of the 10~$\mu$m silicate feature in the quasar absorption systems studied. In addition, we also detect the 18~$\mu$m silicate feature in the sources with adequate spectral coverage. We find variations in the breadth, peak wavelength, and substructure of the 10~$\mu$m interstellar silicate absorption features among the absorbers. This suggests that the silicate dust grain properties in these distant galaxies may differ relative to one another, and relative to those in the Milky Way. We also find suggestions in several sources, based on comparisons with laboratory-derived profiles from the literature, that the silicate dust grains may be significantly more crystalline than those in the amorphous Milky Way ISM. This is particularly evident in the z=0.89 absorber toward the quasar PKS 1830-211, where substructure near 10~$\mu$m is consistent with a crystalline olivine composition. If confirmed, these grain property variations may have implications for both dust and galaxy evolution over the past 9 Gyrs, and for the commonly-made assumption that high-redshift dust is similar to local dust. We also discuss indications of trends between silicate dust absorption strength and both carbonaceous dust properties and gas-phase metal properties, such as the gas velocity spread, determined from UV/optical spectra.}

\FullConference{The Life Cycle of Dust in the Universe: Observations, Theory, and Laboratory Experiments - LCDU 2013,\\
		18-22 November 2013\\
		Taipei, Taiwan}

\begin{document}

\section{Introduction}
Quasar absorption systems (QASs), galaxies seen in absorption along the sightlines to background quasars, are a valuable probe of the gas and dust conditions in the distant universe. Since QASs sample galaxies
by gas cross-section, independent of the galaxy brightness, they can provide a view of normal galaxies at high redshift, which can be missed in flux-limited surveys biased toward brighter or more actively star-forming galaxies. Damped Lyman-$\alpha$
absorbers (DLAs; neutral hydrogen column densities $N_{\rm{H I}} \geq 2\times10^{20}$cm$^{-2}$) and sub-DLAs ($10^{19}\leq N_{\rm{H I}} < 2\times10^{20}$cm$^{-2}$) are particularly important because they provide the primary neutral gas reservoir
for star formation (e.g., [1]). The UV and optical absorption lines produced by metals in QASs yield insight into a range of gas properties, including elemental abundances, kinematics, temperatures, densities, and
ionization parameters. Dust grain properties in QASs can be studied using spectra which cover the 2175~\AA~bump, associated with carbonaceous dust, and the 10 and 18~$\mu$m absorption features associated with silicate dust grains, 
typically produced in oxygen-rich environments. 

Evidence for dust in QASs stems primarily from observed depletions of refractory elements and reddening of background quasars (e.g., [2,3]). A correlation between reddening and metal absorption line strengths yields evidence
of dust in at least some QASs at $1<z<2$ [4]. Most previous studies of dust in QASs have focused on dust signatures associated with the carbonaceous dust, such as the 2175~\AA~bump, rather than on the silicate dust. 
However, in the Milky Way, models find $\sim$66\% of the core mass of interstellar dust grains is silicate in composition (e.g., [5]). 
In the initial study of the z=0.524 DLA toward the quasar AO 0235+164, our team made a 15$\sigma$ detection (in equivalent width; EW) of the 10~$\mu$m silicate absorption feature,
and determined that the shape of the absorption feature was best-matched by profiles for diffuse Galactic interstellar clouds or laboratory amorphous olivine [6], with a
shallow peak optical depth ($\tau_{10}=0.09$). The success of this detection led to 2 programs to study the silicate dust grain properties in 12 additional QASs at z$_{abs}<$1.4 [7,8,9]. 
We report here on the results of these programs, and examine correlations between the silicate absorption strength and other QAS dust and gas properties.

\section{Sample Selection and Data}
The QASs in our samples were chosen to be abundant in both gas and dust, with a single, dominant absorption system along the line of sight and expected $N_{\rm{H I}}>10^{20}$ cm$^{-2}$. 
The full absorber sample is comprised of 13 QASs with $0.156\leq z_{abs} \leq 1.388$. 
All of the systems show multiple signatures of dust such as high extinctions (A$_V>$0.4), reddening of the quasar, detection of the
2175~\AA~bump, significant depletions of refractory elements (e.g., Fe, Cr), and/or the presence of diffuse interstellar bands (DIBs). Some of the absorbers also show evidence of 21 cm and/or
X-ray absorption. The absorbers toward the gravitationally lensed PKS 1830-211 and TXS 0218+357 quasars are among the $\sim$5 known distant QASs displaying strong molecular absorption.

The primary data consist of spectra obtained with the Spitzer Infrared Spectrograph (IRS) in low-resolution mode (SL, LL). The observations were designed to cover the full, rest-frame 10~$\mu$m 
silicate absorption feature in \textit{every} QAS, as well as all or part of the rest-frame 18~$\mu$m silicate absorption feature in 6 of the QASs. These data were processed using the standard IRS reduction pipeline, followed
by applications of the Spitzer {\tt IRSCLEAN} and IRAF {\tt epix} algorithms to identify and replace bad pixels. The 1-D spectra were extracted using the optimal, point-source mode of the Spitzer {\tt SPICE} algorithm. 
The quasar spectrum was continuum-normalized and shifted to the QAS rest-frame, prior to fitting the observed QAS silicate absorption features with template profiles obtained from the literature.
We have also utilized information on the 2175~\AA~bump and metal absorption properties, where available, using SDSS spectra or published values.

\section{Silicate Dust Absorption in QASs}
In each of the sample QASs that we have analyzed as part of our ongoing program [6,7,8,9], we find a $\gtrsim 4\sigma$ detection of the 10~$\mu$m silicate absorption feature. When template profiles for laboratory amorphous olivine, 
a Galactic molecular cloud (based on observations of the Orion Trapezium region), a Galactic diffuse cloud (based on observations of $\mu$ Cep), and a sightline toward the Galactic center (GCS3) are compared with the 10~$\mu$m
absorption profiles for the QASs toward the quasars AO 0235+164, 3C196, Q0852+3435, Q0937+5628, and Q1203+0634, the best match is generally produced by the laboratory amorphous olivine template [7]. However, 
a comparison of the shape and peak wavelength of the 10~$\mu$m absorption features reveals substantial system-to-system variations, as discussed further in Section~\ref{vars}. 

\subsection{z=0.89 QAS toward the Quasar PKS 1830-211}
\begin{figure} 
\center
\includegraphics[width=.24\textwidth]{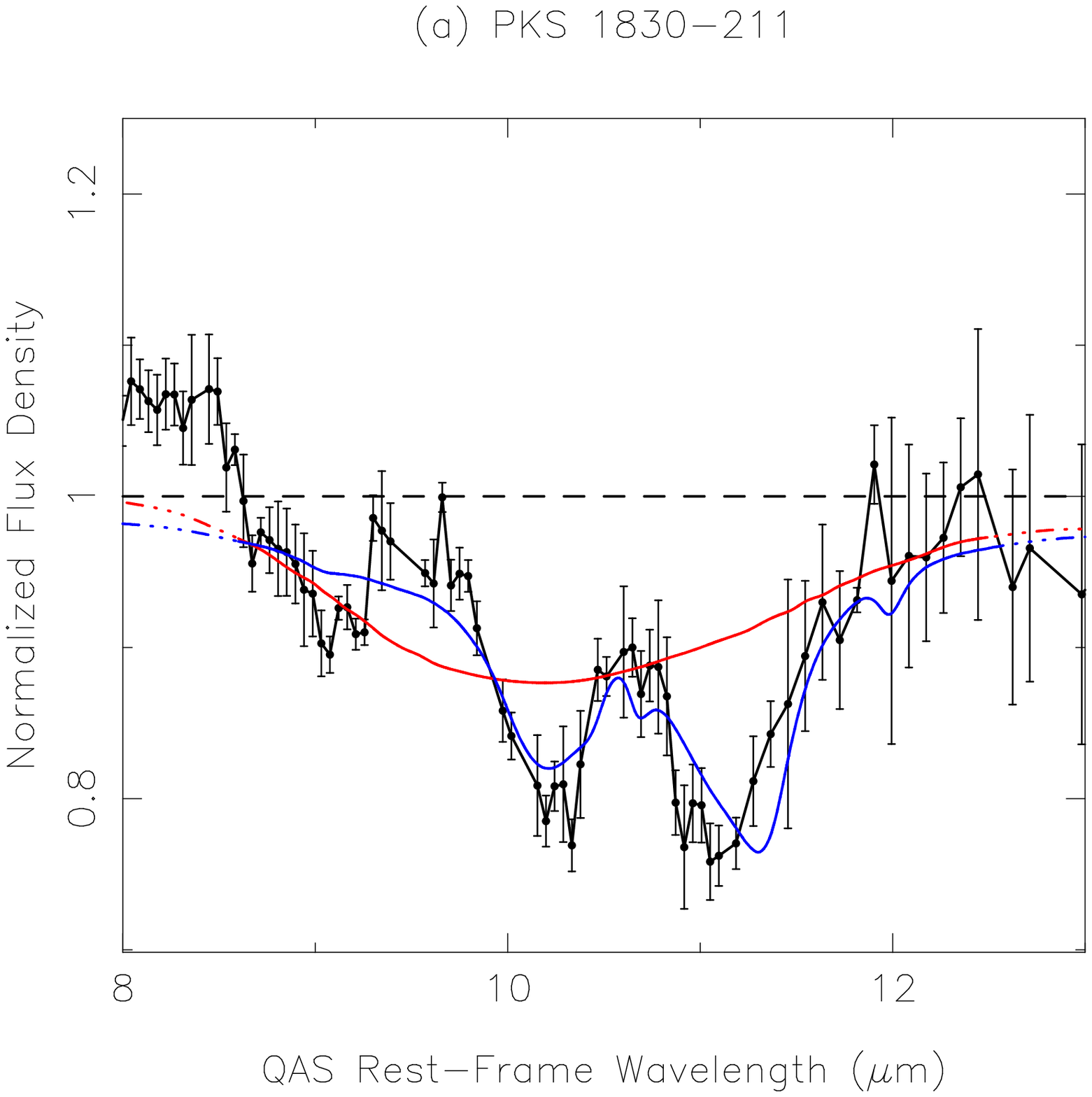} 
\includegraphics[width=.24\textwidth]{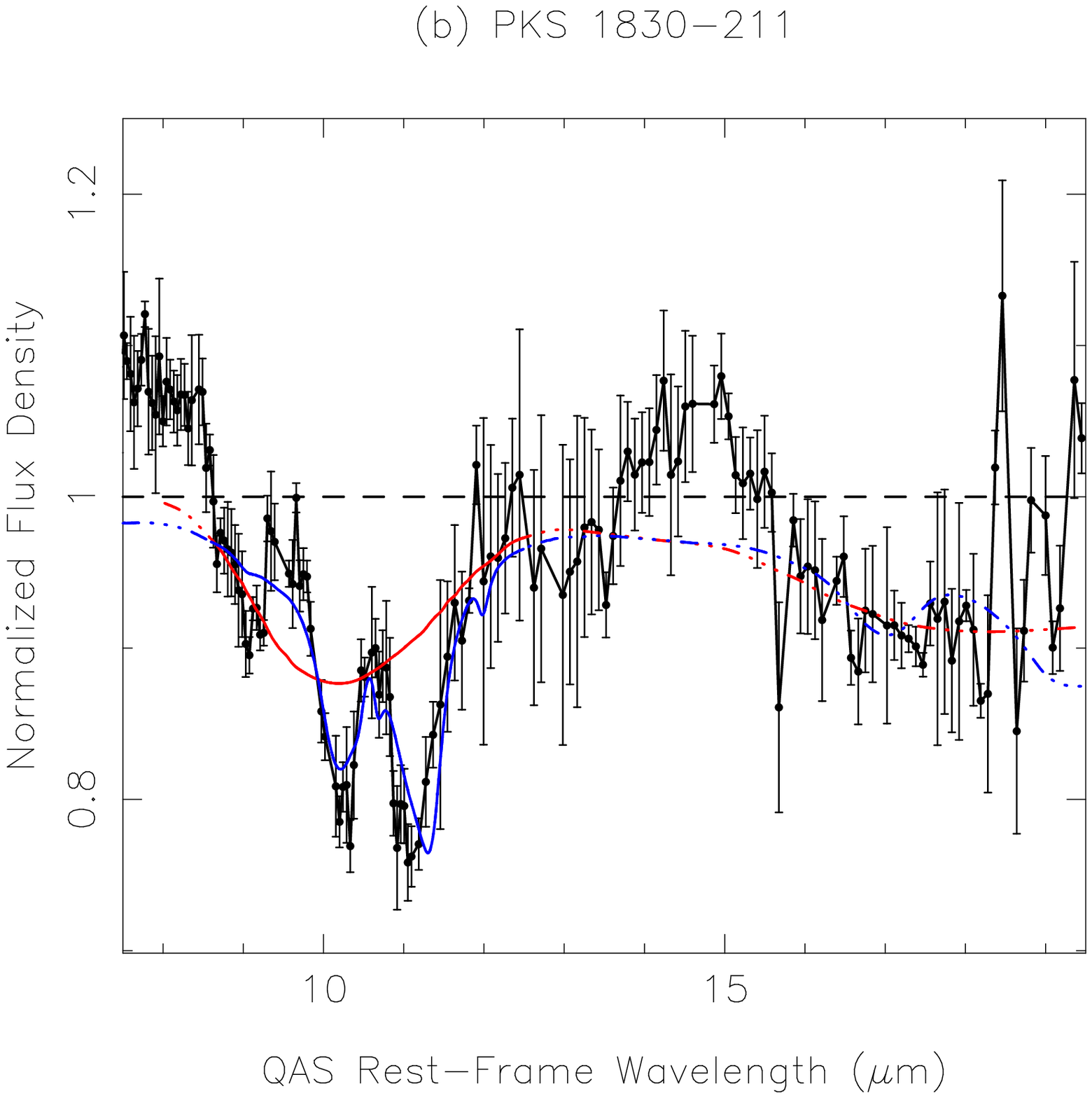} 
\includegraphics[width=.24\textwidth]{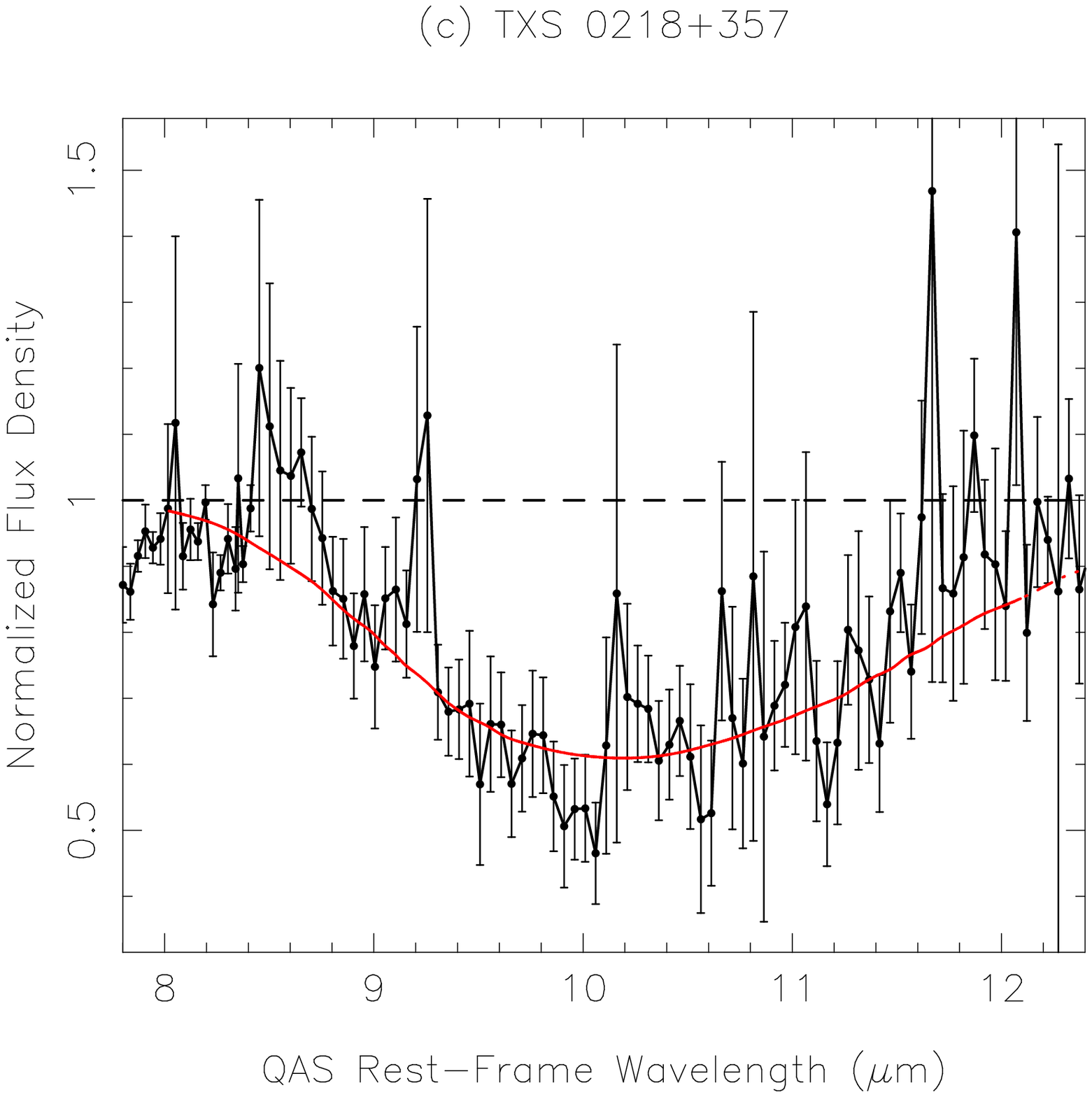}
\includegraphics[width=.24\textwidth]{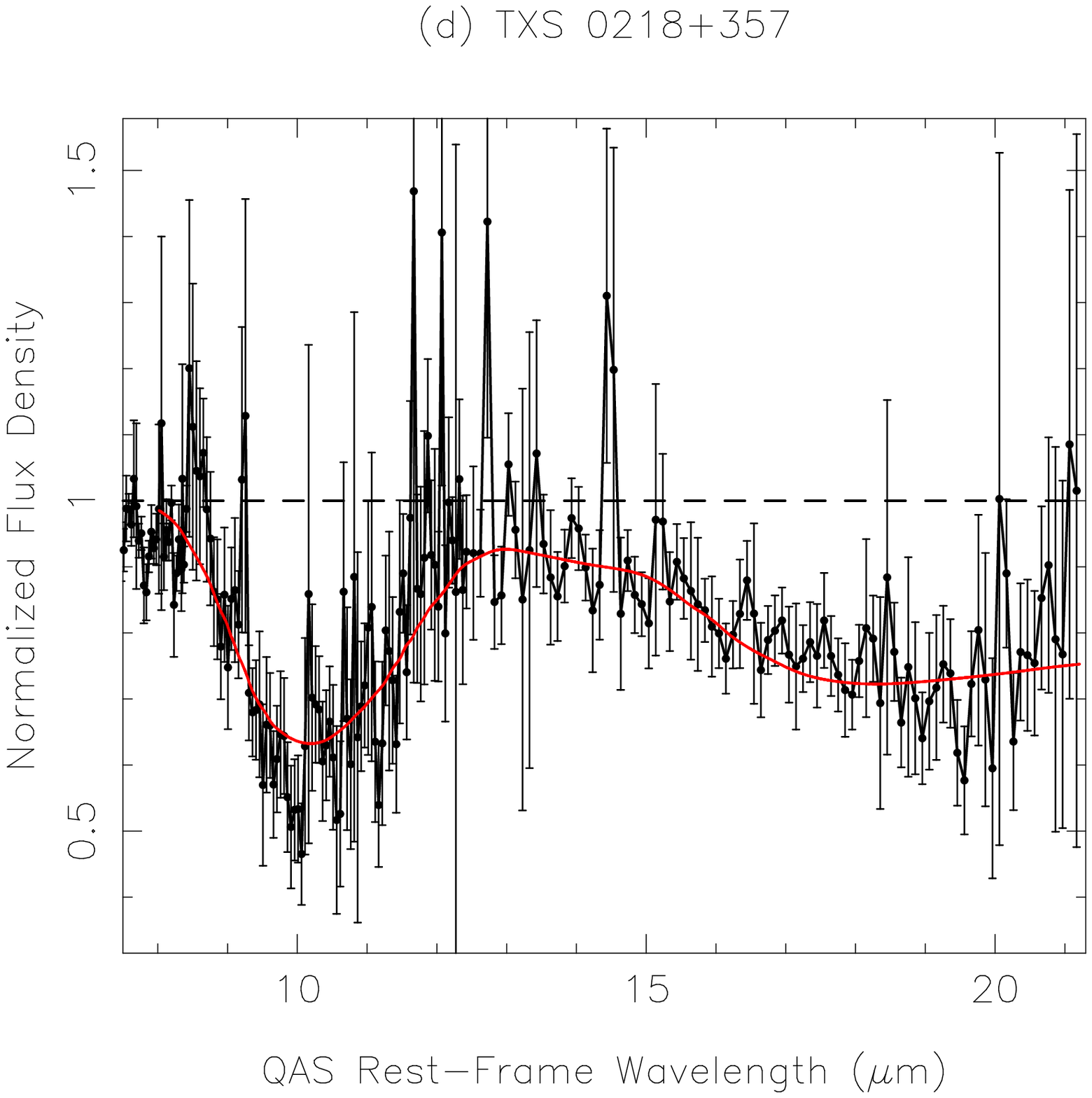}
\caption{Comparison of template profile fits for the z=0.89 QAS toward PKS 1830-211 (a-b) and the z=0.69 QAS toward TXS 0218+357 (c-d). 
(a) Fit to the 10~$\mu$m PKS 1830-211 absorption feature obtained using laboratory amorphous olivine (red; profile with porous, ellipsoidal (PE) particles [13]) 
and crystalline olivine (blue; hortonolite [10]) templates and (b) fit prediction over the 18~$\mu$m region. 
Fits to the TXS 0218+357 (c) 10~$\mu$m and (d) combined 10 and 18~$\mu$m absorption features obtained using the same (PE) laboratory amorphous olivine template (red, [13]).
Fits are depicted with (colored) solid lines over the regions where fitting was performed, and as broken lines outside of the fitting regions.} 
\label{fig1} 
\end{figure}

The system with the most significant 10~$\mu$m region deviations relative to the laboratory amorphous silicate template is the z=0.89 QAS located in a face-on spiral galaxy toward the lensed, background quasar PKS 1830-211.
This QAS is noted to exhibit both 21-cm (H~I) absorption, as well as absorption from more than 25 different molecular species (e.g., CO, HCO$^+$, HCN, H$_2$O, NH$_3$, see, e.g., [10]). 
We detect the 10~$\mu$m silicate absorption feature at 12.7$\sigma$ (in EW), and also find evidence of 18~$\mu$m absorption, which is near the edge of the spectral region covered by our data [8]. 
As illustrated in Figure~\ref{fig1}(a-b), laboratory amorphous olivine provides a poor fit to the observed 10~$\mu$m absorption feature, and does not match the peak absorption wavelength, the breadth
of the feature, or the observed substructure within the broad absorption feature. In order to identify a better match to the observed absorption feature, we explored fits using $>$100 optical depth templates drawn from the literature for both 
astrophysical and laboratory sources. We found that the observed feature was best-matched by laboratory crystalline olivine (Mg$_{2x}$Fe$_{2-2x}$SiO$_4$) profiles, in particular
hortonolite (Mg$_{1.1}$Fe$_{0.9}$SiO$_4$, [11]), with a corresponding peak optical depth of $\tau_{10}=0.27\pm0.01$. Our finding that a crystalline silicate template best matches the QAS interstellar silicate
dust absorption is highly unusual when contrasted with the predominately amorphous Galactic interstellar silicate dust [12]. We explored the possibility that the substructure in the absorption feature we are
attributing to crystallinity could have a different origin including other atomic or molecular (e.g., H$_2$, [S IV], [Ar III]) or PAH features in either the QAS or in the background quasar, the presence
of a second (foreground) QAS along the quasar sightline, or systematic effects associated with the instrument or the spectral extraction software. None of these alternative explanations could explain the substructure
as well as the crystalline silicate absorption profile [8]. If the interstellar silicate dust in this QAS is indeed crystalline, it may suggest a highly unusual environment, possibly also explaining the large diversity of molecules. Future data covering the
20-69~$\mu$m crystalline resonance features would validate the identification of silicate crystallinity.

\subsection{z=0.69 QAS toward the Quasar TXS 0218+357}
In order to further explore whether QASs with strong molecular absorption also exhibit silicate dust crystallinity, we examined the z=0.69 QAS toward the lensed quasar TXS 0218+357. Similar to the z=0.89 PKS 1830-211 QAS,
this absorber is in a face-on spiral galaxy, with absorption from $>$10 different species of molecules including CO, HCO$^+$, H$_2$O, NH$_3$, and LiH, as well as 21-cm absorption. We detect both the 10~$\mu$m silicate absorption feature
at 10.7$\sigma$ (in EW), and additionally detect the 18~$\mu$m feature at $>$3$\sigma$ [9]. 
We do not find
compelling evidence of silicate crystallinity, although, due to the low signal-to-noise ratio (S/N) of the TXS 0218+357 spectrum, we cannot rule out crystalline silicates producing some of the smaller substructure features 
in the absorption profile. 
As illustrated in Figure 1 (c-d), the location and breadth of the 10 and 18~$\mu$m silicate absorption features are well-matched by an amorphous olivine silicate
template, with an inferred peak optical depth of $\tau_{10}=0.49\pm0.02$ when fitting solely over the 10~$\mu$m feature [9]. This is the highest $\tau_{10}$ peak optical depth measured for the QASs we have studied. 

\subsection{Variations in 10~$\mu$m Silicate Feature} \label{vars}
When we compare the 10~$\mu$m silicate absorption feature in the sample QASs (Figure 2a), we see variations in the depth, breadth, and peak absorption wavelength of the feature. We also see variations in 
the absorption feature substructure. While some smaller substructure variations may result from low S/N data, others (e.g., in the PKS 1830-211 spectrum) appear to be intrinsic to the QAS. 
These cumulative differences between the absorption features in the QASs may suggest variations in the silicate dust grain properties (such as the composition and the grain temperatures, sizes, shapes, and porosities).
However, we do not find any correlations in these silicate absorption feature variations with QAS parameters such as the presence of molecules or the absorber redshift.

\begin{figure} 
\center
\includegraphics[width=.235\textwidth]{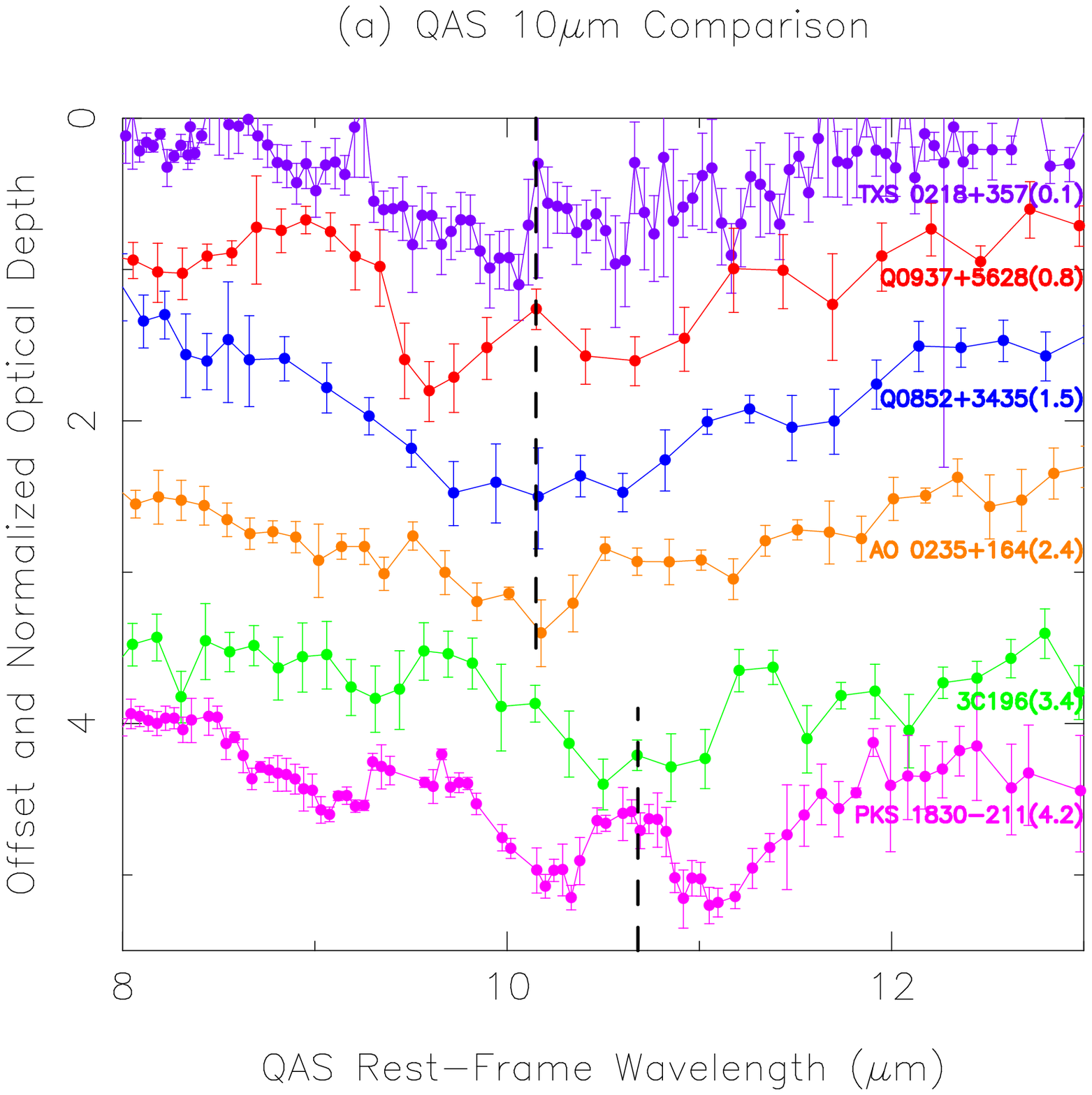} 
\includegraphics[width=.24\textwidth]{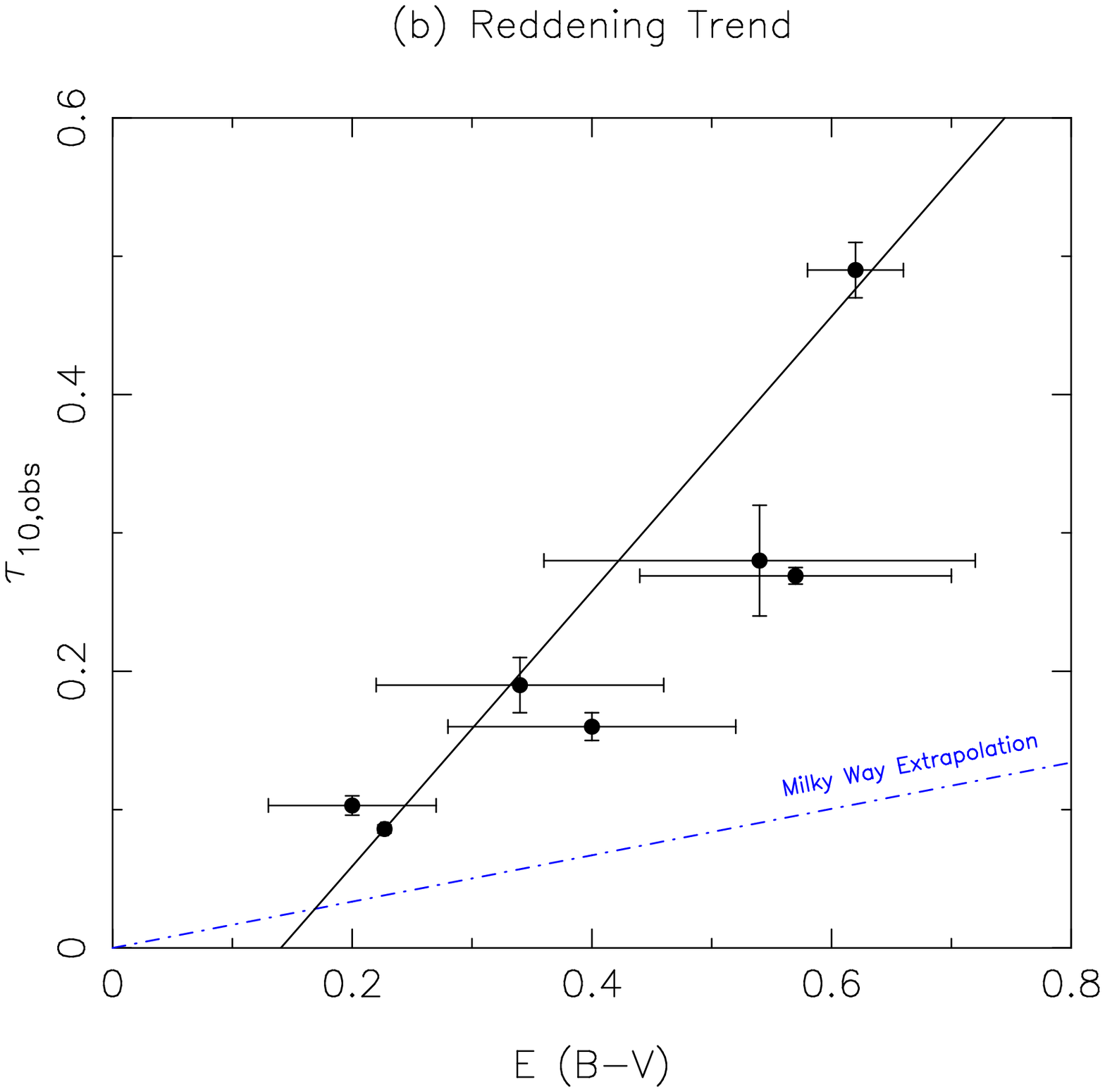} 
\includegraphics[width=.24\textwidth]{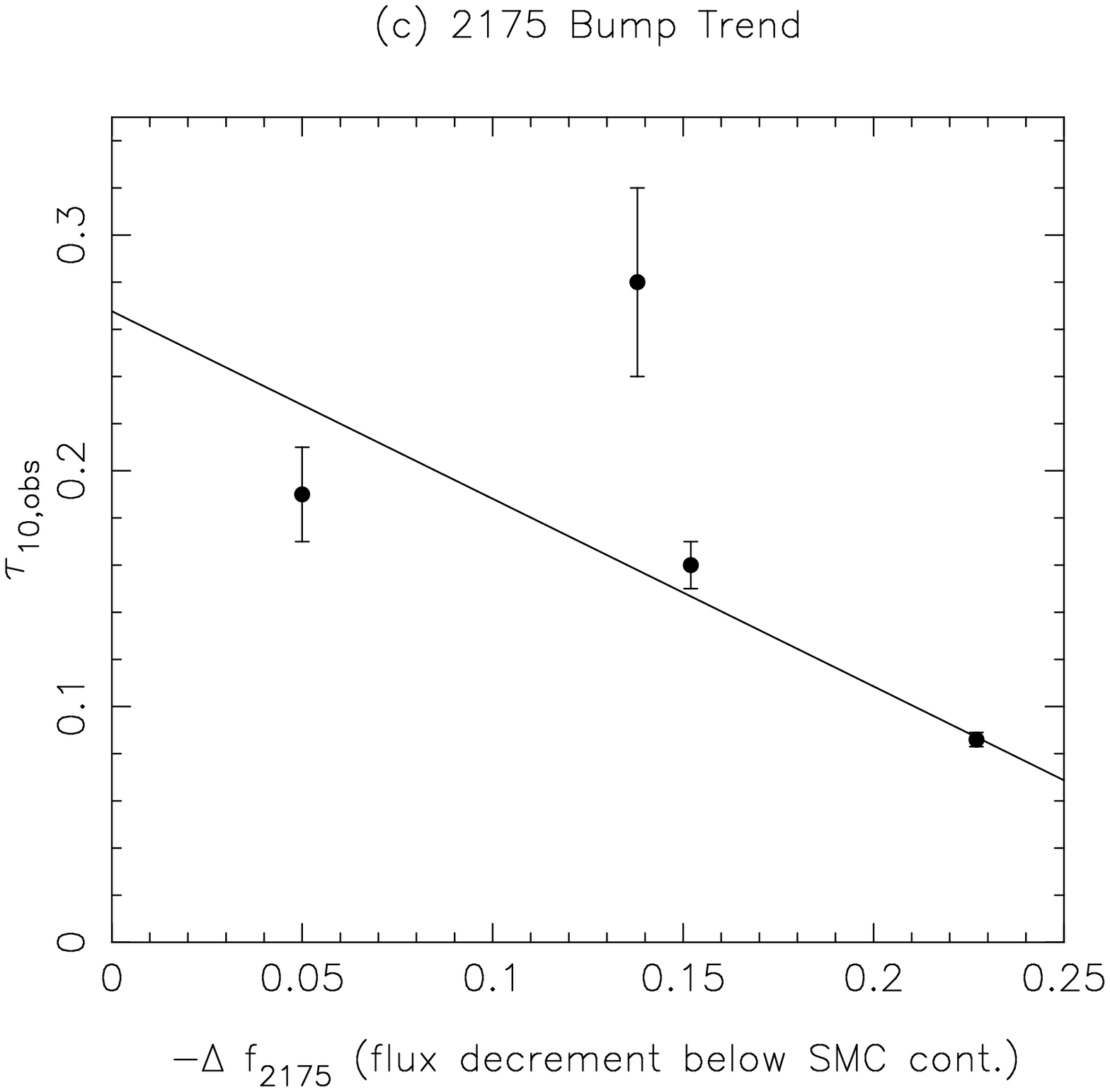} 
\includegraphics[width=.235\textwidth]{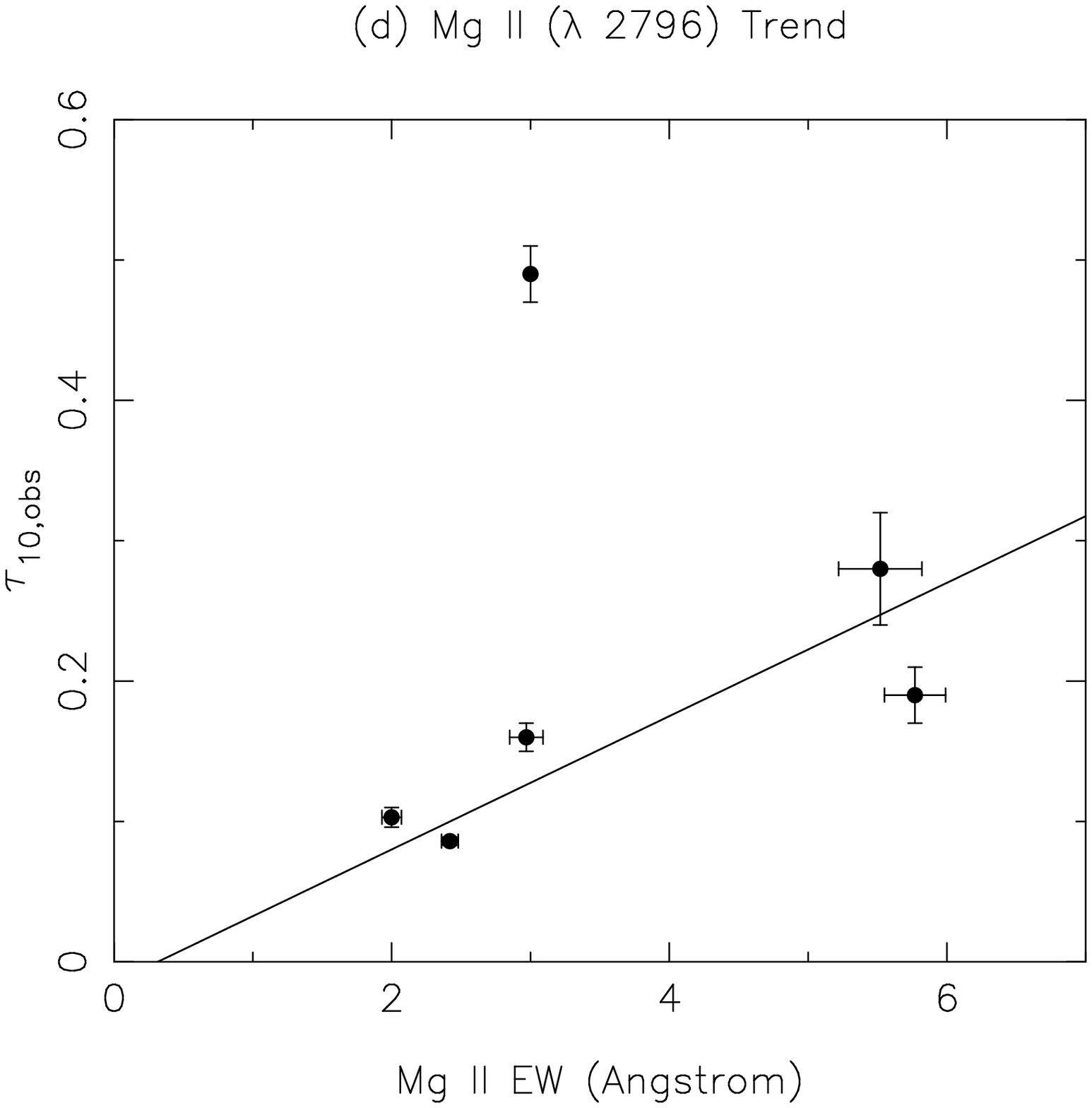}
\caption{(a) Comparison of the normalized 10~$\mu$m optical depth profiles for 6 QASs [7,8,9], illustrating variations in the feature shape and breadth. 
The profiles are labelled by the name of the quasar, and vertically offset for visual clarity (with offset given in parenthesis). The dashed vertical lines emphasize the differences
in peak absorption wavelength between the lower and upper QASs in the panel. 
Examination of trends between the 10~$\mu$m peak optical depth and (b) the reddening, (c) the strength of the 2175~\AA~bump (defined as the flux decrement below SMC continuum),
and (d) the Mg~II $\lambda$2796 rest-frame EW [7,9]. The black line shows the best fit to the data in each panel, with a covering factor of unity assumed
for all of the QASs.} 
\label{fig2} 
\end{figure}

\section{Silicate Dust Correlations with Other Dust and Gas Properties}
In Milky Way diffuse clouds there is an established correlation between the reddening, $E(B-V)$, and the peak optical depth of the 10~$\mu$m silicate absorption feature, $\tau_{10}$, with a slope of 0.17 
for R$_V$=3.1 [14]. Our sample QASs also show a correlation between $E(B-V)$ and $\tau_{10}$, but the slope of the trend is $\sim$3-6$\times$ higher than the extrapolation from the Galactic diffuse clouds, 
depending on which QASs are included in the fit [9]. As depicted in Figure 2b, the largest offset relative to the Galactic trend occurs for the TXS 0218+357 QAS, which has a high $\tau_{10}$ for the measured $E(B-V)$; if 
covering factors $<1$ are utilized, this offset becomes even more substantial. The reason for the steeper QAS trend relative to that in the Milky Way may stem from grain differences. 
For instance, larger dust grains could result in a lower UV extinction. Alternatively it is possible that in face-on systems, such as the PKS 1830-211 and TXS 0218+357 QASs (which exhibit a much higher $\tau_{10}$ than predicted from their
$E(B-V)$ if similar to the Milky Way diffuse clouds), a different or more limited grain population is being sampled than along sightlines integrated through the Galactic disk. Another possibility is that there are different stellar populations in these QASs, 
for instance more O-rich stars, which would boost the amount of silicate dust relative to Galactic sightlines. 

We have also explored trends between silicate and carbonaceous dust absorption strengths in 4 QASs with data covering the 2175~\AA~bump [7]. As illustrated in Figure 2c, 
we find a slight hint of an anti-correlation between the carbonaceous and silicate dust absorption strengths. This may suggest that dust is either predominately carbonaceous or predominately silicate
along a sightline through the absorber, although data for more systems are required to investigate further.

In addition to exploring trends with other dust properties, we have investigated a trend between the silicate dust absorption strength ($\tau_{10}$)
and the Mg~II $\lambda$2796 absorption line rest-frame EW [9], as illustrated in Figure 2d. In most of these
systems Mg~II is saturated, making the equivalent width a proxy for the velocity spread (hence related to outflow speeds or the dynamical speed), rather than the elemental abundance. 
The observed trend may, therefore, be implying that silicate-rich QASs are more massive. We note
that the TXS 0218+357 QAS is an outlier in the relation, possibly because of differences in saturation [9].

\section{Summary and Conclusions}
In summary, we find 10~$\mu$m (and 18~$\mu$m) silicate absorption in each of the investigated dusty, z$<$1.4 QASs. However, observed variations in the shape,
breadth, and location of the absorption features suggest that there may be differences in the silicate dust grain properties between the systems. We also find evidence of trends between
the silicate absorption strength ($\tau_{10}$) and other QAS dust and gas properties. In particular, we find a correlation with the reddening which is steeper than found
for Galactic diffuse clouds, a correlation with the Mg~II EW which may suggest that silicate-rich QASs are more massive, and a mild suggestion of an anti-correlation with the carbonaceous dust
absorption strength. In the immediate future, we will be exploring these suggested trends using an expanded sample of QASs with archival dust and gas data, 
and examining the interrelation between gas and dust properties in connection with dust/chemical evolution models for galaxies. 

\medskip

\noindent \textbf{Acknowledgements.} Support in the form of an International Travel Grant from the American Astronomical Society and the US National Science Foundation (NSF) to MCA enabled attendance at this conference.
Support for this work is provided by NASA through an award issued by JPL/Caltech and from NSF grants AST-0908890 and AST-1108830 to the U. of South Carolina.

\end{document}